\documentclass{article}

\usepackage{booktabs}
\usepackage{amsmath,amsthm,amsfonts,amssymb,mathrsfs,mathtools,bm}
\usepackage{graphicx,color,subcaption,float,soul,multirow,makecell,blkarray}
\usepackage{algpseudocode}
\usepackage{xparse}
\usepackage{hyperref}
\usepackage{enumerate}
\usepackage[font={footnotesize,bf},justification=centering]{caption}
\usepackage{authblk,xspace}
\usepackage{framed}
\usepackage{xspace}
\usepackage{relsize}
\usepackage{textcomp}
\usepackage[english]{babel}
\usepackage{pifont}

\usepackage{tikz}
\usetikzlibrary{calc,positioning,shadows,shapes,mindmap}

\newtheorem{example}{Example}
\newtheorem{definition}{Definition}

\newtheorem{remark}{Remark}

\newcommand{\RR}{\ensuremath{\mathbb{R}}\xspace}

\newcommand{\Bc}{\ensuremath{\mathcal{B}}\xspace}

\newcommand{\Nc}{\ensuremath{\mathcal{N}}\xspace}

\newcommand{\Pc}{\ensuremath{\mathcal{P}}\xspace}

\newcommand{\Tc}{\ensuremath{\mathcal{T}}\xspace}

\NewDocumentCommand{\Xv}{O{n}O{1}}{x_{#2},\dots,x_{#1}}

\NewDocumentCommand{\Tv}{O{d}O{1}}{t_{#2},\dots,t_{#1}}

\newcommand*{\diff}{\mathop{}\!\mathrm{d}}

\DeclareMathOperator{\Iso}{Iso}

\DeclareMathOperator{\Id}{Id}

\DeclareMathOperator{\Vol}{Vol}

\DeclareMathOperator{\Bound}{\text{\raise1pt\hbox{$B$}\hspace{-7pt}\hbox{$B$}}}
\DeclareMathOperator{\Translation}{Transl}
\DeclareMathOperator{\Center}{Centre}
\DeclareMathOperator{\Rotation}{Rot}
\DeclareMathOperator{\Dist}{Dist}
\DeclareMathOperator{\Diameter}{Diameter}
\DeclareMathOperator{\Cost}{Cost}

\newcolumntype{L}{>{$}l<{$}}
\newcolumntype{C}{>{$}c<{$}}
\newcolumntype{R}{>{$}r<{$}}

\newcommand{\setor}{\text{ or }}
\newcommand{\setand}{\text{ and }}
\newcommand{\setst}{\text{ such that }}

\DeclarePairedDelimiter{\abs}{\lvert}{\rvert}
\DeclarePairedDelimiter{\norm}{\lVert}{\rVert}

\newcommand\quotient[2]{
    \mathchoice{
        \text{\raise1ex\hbox{$#1$}\big/\lower1ex\hbox{$#2$}}%
    }{
        #1\,/\,#2
    }{
        #1\,/\,#2
    }{
        #1\,/\,#2
    }
}

\makeatletter
\newcommand{\labeltext}[2]{%
  \@bsphack
  \csname phantomsection\endcsname 
  \def\@currentlabel{#1}{\label{#2}}%
  \@esphack
}
\makeatother

\colorlet{shade}{gray!5}

\algrenewcommand{\Return}{\State\algorithmicreturn~}
\algnewcommand{\IIf}[2]{\State\algorithmicif\ #1\ \algorithmicthen\ #2}
\newcommand{\Continue}{\State\textbf{continue}}
\algrenewcommand{\algorithmiccomment}[1]{\null\hfill(\textit{#1})}
\algnewcommand{\algorithmiccommenttwo}[2]{\null\hfill(\textit{#1}\linebreak[4]\null\hfill\textit{#2})}
\algnewcommand{\CommentTwo}[2]{\algorithmiccommenttwo{#1}{#2}}

\title{An Implicit Representation of Swept Volumes based on Local Shapes and Movements\\
{\small Technical Report}}
\author{Cl{\'e}ment Laroche}

\begin{document}

\maketitle

\section{Introduction}\label{Sec:Introduction}

We introduce a new way to implicitly represent swept volumes in 3D.
\emph{Swept volumes} are the trace of objects that are swept along a \emph{rigid transformation}, that is a non-distorting smooth transformation.
Swept volumes are used in \hyperref[Abbr:CAGD]{CAGD} for designing 3D objects through boolean operations or in robotics for searching a path avoiding obstacles.
Considering a 3D object $\Bc$ used as a shaping tool, like a drilling or milling machine, that progressively removes parts of a 3D object $O$, the movement of the tool follows a time-dependent rigid transformation $\Tc(t)$ provided that this shaping tool cannot be deformed.
The result of this operation is a shaped 3D object $O'$ that is the difference of the base object by the tool swept along that rigid transformation:
\[
O' = O \backslash \Tc(\Bc), \text{ where } \Tc(\Bc) = \cup_t [\Tc(t)](\Bc)
\]

The goal of this report is to give an efficient implicit representation of the swept volume $\Tc(\Bc)$.
This implicit representation is then used to perform the above boolean difference with the object to be shaped.
It can also be used for collision detection in problems where the goal is for obstacles $O$ and object paths $\Tc(\Bc)$ to avoid collision.

In the following, $\Bc$ is called the \emph{base volume} in opposition to the swept volume $\Tc(\Bc)$.

Starting with a point cloud of the base volume, we build a data structure enabling this kind of operations:
\begin{itemize}\labeltext{implicit operations}{List:ImplicitOperations}
\item Given a point $P\in\mathbb R^3$, does the point $P$ belong to the swept volume $\Tc(\Bc)$?
\item What is the distance between $P\in\mathbb R^3$ and $\Tc(\Bc)$?
\item Given a ray $R$, what is the first intersection of $R$ with $\Tc(\Bc)$? What are all its intersections?
\item Given an other object $O$, what is the boolean subtraction $O \backslash \Tc(\Bc)$?
\end{itemize}

\begin{figure}[tb]
\hspace{-65pt}
\begin{tikzpicture}[mindmap,
	->, >=stealth,
	every concept/.style={
		rectangle, rounded corners, minimum height=10pt, minimum width=20pt, align=center, draw=black, text width=70pt
	},
	every node/.style={
		font=\fontsize{9}{10}\selectfont,
		inner sep=3pt
	}, x=1pt, y=1pt]

\coordinate (A) at (-40,135);
\coordinate (B) at (40,15);

\draw[drop shadow,rounded corners,fill=blue!20] (A) rectangle (B);

\node[font=\fontsize{11}{12.5}\selectfont] (InputMsg) at (0,125) {Inputs:};
	\node[concept, concept color=green!50] (InputPC) at (0,95) {Point \\ Cloud};
	\node[concept, concept color=green!50] (InputTransf) at (0,30) {Rigid \\ Transformation};

\node[concept, concept color=purple!40] (RiscPC) at (-100,75) {
	Point Cloud of swept volume\\
	\includegraphics[width=70pt]{SweptPointCloud.png}
};
\node[concept, concept color=purple!40] (RiscImpl) at (-200,75) {
	Implicit \\ representation of swept volume\\
	\includegraphics[width=70pt]{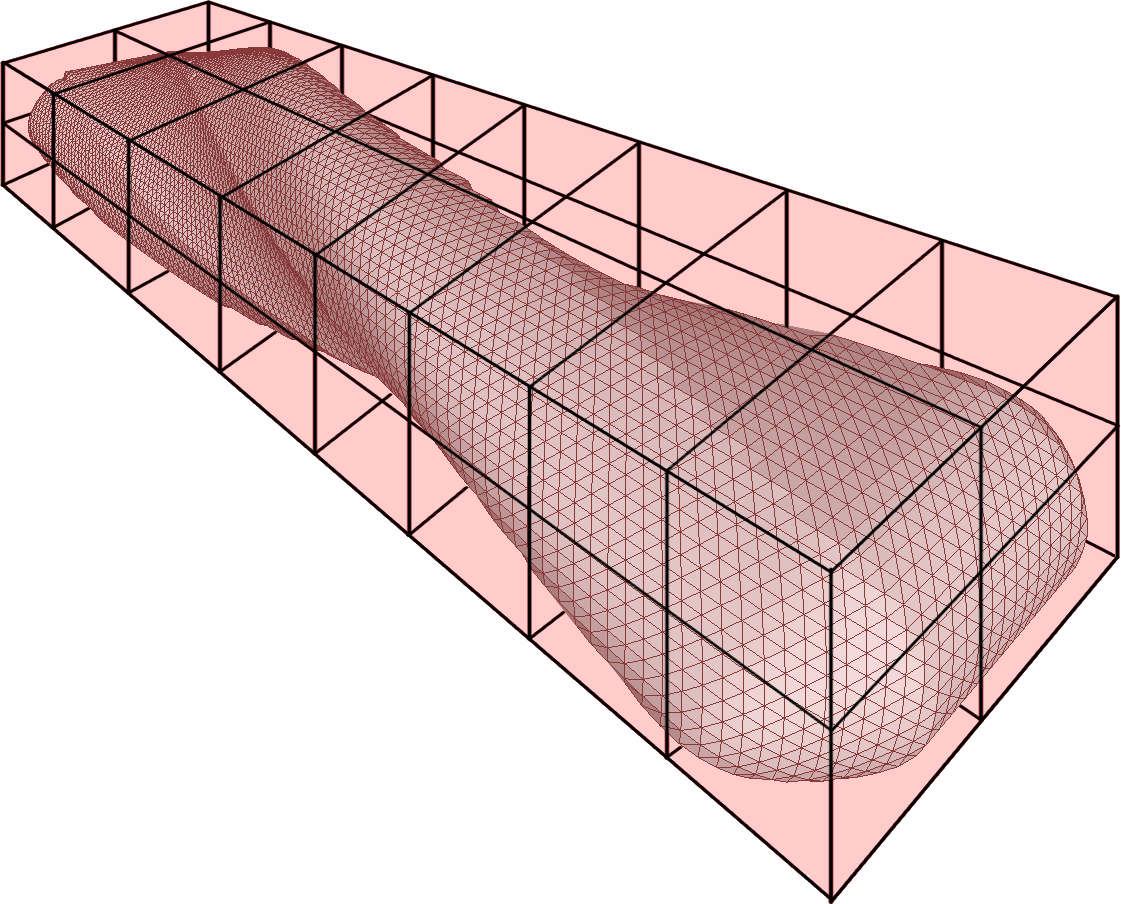}
};

\node[concept, concept color=gray!60] (Base) at (100,100) {
	Implicit \\ representation of base volume\\
	\includegraphics[width=70pt]{LRBSplineTool.png}
};
\node[concept, concept color=gray!60] (SweptVolume) at (190,50) {
	Implicit \\ representation of swept volume\\
	\includegraphics[width=70pt]{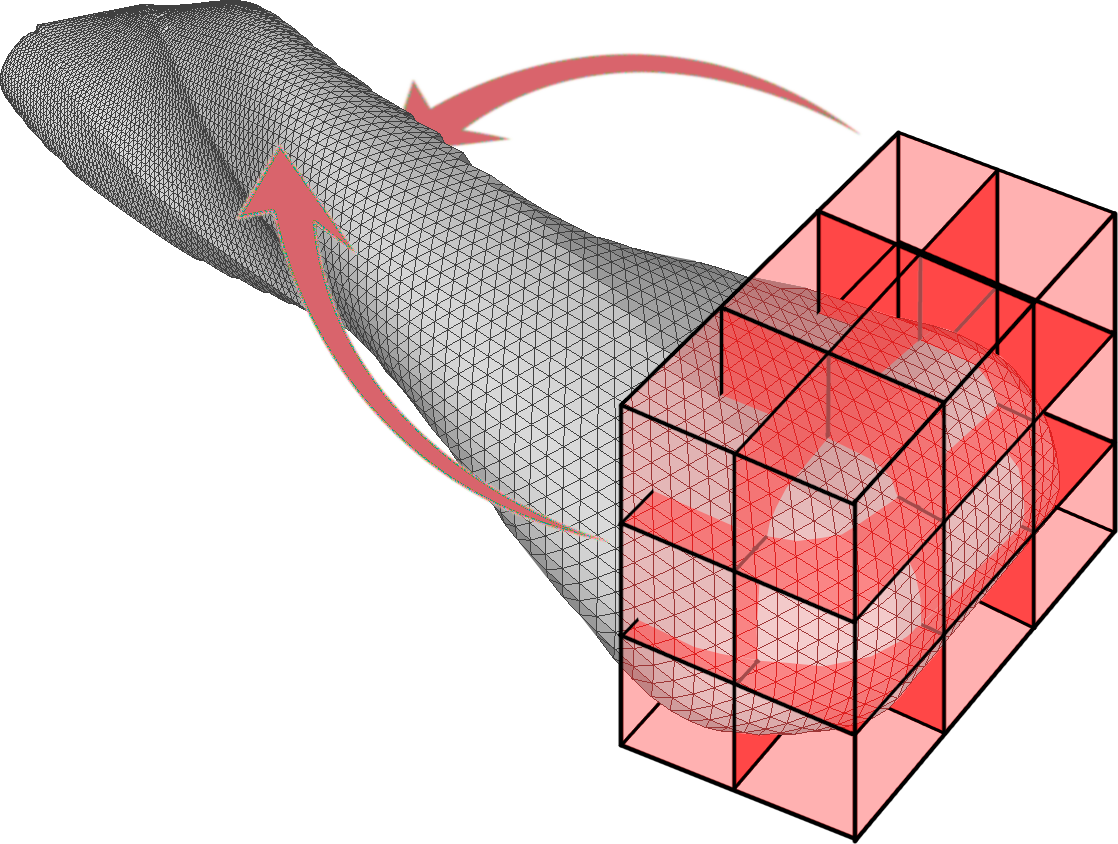}
};

\draw[thick] (InputPC.south) to[bend left=10] (RiscPC);
\draw[thick] (InputTransf.north) to[bend right=8] (RiscPC);
\draw[thick] (RiscPC) to (RiscImpl);
\draw[thick] (InputPC) to (Base);
\draw[thick] (InputTransf) to[bend right=10] ($(SweptVolume.south west)+(0, 20)$);
\draw[thick] (Base) edge[-, bend right=30] ($(SweptVolume.south west)+(-30, 15.2)$);
\end{tikzpicture}
\caption{Two possible representation strategies: construct a data structure of the swept point cloud (\emph{left}), construct a data structure using both the point cloud of the base volume and transformation informations (\emph{right})}
\label{Fig:SweptRepAlternatives}
\end{figure}

One way to proceed is first generate a point cloud of the swept volume and then implicitize that point cloud 
(this strategy is depicted on the left part of Fig.~\ref{Fig:SweptRepAlternatives}).
Here, however, we take a different path: we first implicitize the base volume and only then 
we use the transformations to build an implicit representation of the swept volume 
(depicted on the right part of Fig.~\ref{Fig:SweptRepAlternatives}).
This way, we can build an implicit representation that fits to the swept feature of $\Tc(\Bc)$, 
allowing more details in its geometry due to the fact that 
the details of the base volume is carried to the details of the swept volume.

Constructing a data structure suited for swept volumes not only allows to perform~\ref{List:ImplicitOperations} 
but also give more specific answers such as: 
if a point $P$ belongs to the swept volume, for which $t$ is it inside $[\Tc(t)](\Bc)$?
which part(s) of the base volume meet with $P$? etc.

\medskip

After recalling how an implicit representation of $\Bc$ can be constructed from a point cloud, we present our method to construct an implicit representation structure of $\Tc(\Bc)$ out of that representation of $\Bc$.
This structure is flexible enough for allowing different kinds of representations as input and have a correct balance of geometric details generated by the rigid transformation and details of the base volume $\Bc$ itself.
We then discuss of the usages of this structure and develop points that may be improved in our construction.

\section{Implicitizing a point cloud}\label{Sec:ImplicitizingPC}

\begin{definition}\label{Def:BaseVolume}
A \emph{local implicit representation} of a base volume $\Bc$ is 
a collection $(A_i, F_i)_{1\le i \le N}$ of bounded areas $A_i$ (cubes, balls, \dots)
and of implicit procedures $F_i: A_i \rightarrow \RR$.
The 3D model $\Bc$ is then given by
\[
\Bc = \left\lbrace (x, y, z) \mid (x, y, z) \in A_i \text{ and } F_i(x, y, z) \le 0 \right\rbrace.
\]
\end{definition}

We describe two different ways of constructing a local implicit representation of a base volume from a given point cloud.
The two algorithms presented here need a point cloud with both point coordinates and normals.
These examples of implicitization algorithms exhibit the variety of local implicit representations and thus support the need for the flexible definition given above.

In the following, $\Pc$ and $\Nc$ are a given 3D point cloud of an object's surface and the outer normals of these points, respectively.

\subsection{MPU method}\label{Subsec:MPU}

The Multi-level Partition of Unity implicitization\cite{MPU2003}, or MPU, is an algorithm generating an octree-based local implicit representation.
In other terms, the areas $A_i$ are cuboids whose edges are parallel to the axes.
A local approximation procedure $F_i$ can be of three types in order to adapt to the local shape of $\Bc$:
\begin{enumerate}[(a)]
\item a general 3D quadratic polynomial,\label{Enum:MPUa}
\item a bivariate quadratic polynomial in local coordinates,\label{Enum:MPUb}
\item a piecewise quadratic polynomial for representing edges and corners (2, 3 or 4 pieces depending on the situation).\label{Enum:MPUc}
\end{enumerate}

At each step of the algorithm, we subdivide the cuboids inside which the local approximation are not precise enough into 8 smaller cuboids.
Then we update the local approximations inside these 8 smaller cuboids.
In order to increase the representation's smoothness, the local approximation inside a cuboid is computed by taking into account all the points inside an ellipsoid containing the cuboid.
The precision of a local approximation is computed using the Taubin distance (see~\cite{Taubin1991}).

\begin{figure}
\caption{Algorithm - MPU}
\label{Algo:MPU}
\begin{tabular}{ll}
\textbf{Input}: 	& A point cloud $\Pc$. \\
				& The normals $\Nc$ of that point cloud. \\
\textbf{Output}:	& A suitable local implicit representation for $\Pc$.
\end{tabular}
\begin{algorithmic}[1]
	\State Compute a bounding box $A_1$ of $\Pc$.
	\State Rescale such that $A_1$ is a cube of diagonal length $1$, i.e. of edge length $\frac{1}{\sqrt{3}}$.
	\State Let $A = F = \emptyset$.
	\State Let $S = \{A_1\}$.
	\While{$S$ is not empty}
		\State Pick $a \in S$; $a$ is a cube of diagonal length $d$.
		\State Let $C$ be the sphere centred on the cube $a$ of radius $R=\alpha d$.
		\CommentTwo{typically,}{$\alpha = 0.75$}
		\If{$C$ contains less than $N_{\min}$ points} \Comment{typically, $N_{\min}=15$}
			\State Let $C'$ be an enlargement of $C$ that contains at least $N_{\min}$ points.
		\EndIf
		\State Let $f$ be a MPU local approximation of $\Pc \cap C'$. \Comment{see~\ref{Algo:MPUApprox}}
		\If{it failed}
			\State Subdivide $a$ into $8$ cubes and add them to $S$.
			\Continue
		\ElsIf{$C$ contains no point}
			\State Add $a$ to $A$ and $f$ to $F$.
			\Continue
		\EndIf
		\State Let $\epsilon = \max_{p \in \Pc \cap C} \abs{f(p)}/\norm{\nabla f(p)}$.
		\If{$\epsilon<\epsilon_0$} \Comment{typically, $\epsilon_0 = 10^{-4}$}
			\State Add $a$ to $A$ and $f$ to $F$.
		\Else
			\State Subdivide $a$ into $8$ cubes and add them to $S$.
		\EndIf
	\EndWhile
	\Return $(A, F)$ after rescaling it back.
\end{algorithmic}
\end{figure}

When computing local approximations of type~(\ref{Enum:MPUa}), we first generate a small pointset $Q$ that can be used to obtain a reliable estimate of a signed distance function.
We then compute the quadratic polynomial $f$ that minimizes the following quantity:
\begin{equation}\label{Eq:MPUa}
\frac{1}{\sum_i w(p_i)} \sum_i w(p_i) f(p_i)^2 + \frac{1}{\abs{Q}}\sum_{q\in Q}\left(f(q)-d\right)^2
\end{equation}
where $w, d$ and $Q$ are defined in the algorithm~\ref{Algo:MPUApprox}.

In the case of local approximations of type~(\ref{Enum:MPUb}), a local coordinate system $(u,v,n)$ is introduced, where $n$ is a weighted arithmetic mean of the point cloud's normals.
A bivariate quadratic polynomial $f$ in $(u,v,n)$ is then a polynomial of the form:
\begin{equation}\label{Eq:MPUb}
f(p) = w - \left(c_{20}u^2+c_{11}uv+c_{02}v^2+c_{10}u+c_{01}v+c_{00} \right)
\end{equation}
where $c_{ij}$ are the polynomial's parameters and $u,v,w$ are the coordinates of the point $p$ in the coordinate system $(u,v,n)$.

In the case~(\ref{Enum:MPUc}) of a sharp feature, we compare the different normals in order to determine whether there is an edge, a three-sided corner or a four-sided corner (see~\cite{Kobbelt2001}).
Then, we split the points into two, three or four pointsets respectively and compute local approximations $f_k$ of types~(\ref{Enum:MPUb}) on each of these pointsets separately.
The local implicit procedure is then given by $f(p) = \min_k f_k(p)$.

The algorithm~\ref{Algo:MPU} sketches the main loop of MPU while the algorithm~\ref{Algo:MPUApprox} details the computation of the different types of local approximations.

\begin{figure}
\caption{Algorithm - Local MPU Approximation}
\label{Algo:MPUApprox}
\begin{tabular}{ll}
\textbf{Input}: 	& A sphere of centre $c$ and radius $R$. \\
				& A cube $a$ with the same centre $c$ and diagonal length $d$. \\
				& A point cloud $\Pc'=(p_i)_i$ of at least $N_{\min}$ points inside that sphere. \\
				& The normals $\Nc'=(n_i)_i$ of that point cloud. \\
\textbf{Output}:	& A local approximation of $\Pc'$.
\end{tabular}
\begin{algorithmic}[1]
	\State Let $n = \sum_i b\left(\frac{3\norm{p_i-c}}{2R}\right) n_i$ where $b$ is the quadratic B-Spline then normalise $n$.
	\State Let $\theta$ be the maximal angle between $n$ and $n_i \in \Nc'$.
	\If{$\abs{\Pc'} > 2N_{\min}$ and $\theta\ge\pi/2$} \Comment{case (\ref{Enum:MPUa})}
		\State Let $Q$ be the corners of $a$ and its centre. \Comment{$\abs{Q}=9$}
		\For{$q\in Q$}
			\State Get the 6 nearest neighbours $p^{(j)}$ of $q$ in $\Pc'$. \Comment{$j=1,\dots,6$}
			\IIf{$n^{(j)} \cdot \left(q-p^{(j)}\right)$ have different signs}{Remove $q$ from $Q$.}
		\EndFor
		\IIf{$Q$ is empty}{\algorithmicreturn\ FAIL.}
		\Return $f$ minimizing (\ref{Eq:MPUa}).
	\ElsIf{$\abs{\Pc'} > 2N_{\min}$ and $\theta<\pi/2$} \Comment{case (\ref{Enum:MPUb})}
		\State Let $(u,v,n)$ be an orthonormal local coordinate system centred on $c$.
		\Return $f$ of the form (\ref{Eq:MPUb}) minimizing $\sum_i w(p_i)f(p_i)^2$.
	\Else \Comment{case (\ref{Enum:MPUc})}
		\State Let $p^{(1)}, p^{(2)} \in \Pc'$ and $\theta$ such that $\theta = n^{(1)} \cdot n^{(2)} = \min_{i,j}n_i \cdot n_j$.
		\IIf{$\theta \geq \theta_{\text{sharp}}$}{\algorithmicreturn\ $f$ of the form (\ref{Enum:MPUb})}
		\Comment{typically, $\theta_{\text{sharp}} = 0.9$}
		\State Split $\Pc'=\Pc'_1 \cup \Pc'_2$ using a spherical Voronoi partition w.r.t. $n^{(1)}$ and $n^{(2)}$.
		\Comment{see~\cite{Na2002}}
		\State Let $e = n^{(1)} \times n^{(2)}$, an approximate of the direction of the potential edge.
		\If{$\max_i\abs{n_i \cdot e} \leq \theta_{\text{corner}}$}
		\Comment{typically, $\theta_{\text{corner}} = 0.7$}
			\State Let $f_1, f_2$ of the form (\ref{Enum:MPUb}) w.r.t. $\Pc'_1, \Pc'_2$ respectively.
			\Return $f = \min(f_1, f_2)$.
		\EndIf
		\For{$p_i \in \Pc'$}
			\If{$\abs{n^{(1)} \cdot n_i} < \abs{e \cdot n_i}$ and $\abs{n^{(2)} \cdot n_i} < \abs{e \cdot n_i}$}
				\State Add $p_i$ to a third set $\Pc'_3$ and remove it from $\Pc'_1$ or $\Pc'_2$.
			\EndIf
		\EndFor
		\State Let $p^{(3)}, p^{(4)} \in \Pc'_3$ such that $n^{(3)} \cdot n^{(4)}$ is the smallest amongst points in $\Pc'_3$.
		\If{$n^{(3)} \cdot n^{(4)} \geq \theta_{\text{sharp}}$}
			\State Let $f_1, f_2, f_3$ of the form (\ref{Enum:MPUb}) w.r.t. $\Pc'_1, \Pc'_2, \Pc'_3$ respectively.
			\Return $f = \min(f_1, f_2,f_3)$.
		\EndIf
		\State Split $\Pc'_3 = \Pc'_4 \cup \Pc'_5$ using a spherical Voronoi partition w.r.t. $n^{(3)}$ and $n^{(4)}$.
		\State Let $f_1, f_2, f_4,f_5$ of the form (\ref{Enum:MPUb}) w.r.t. $\Pc'_1, \Pc'_2, \Pc'_4, \Pc'_5$ respectively.
		\Return $f = \min(f_1, f_2,f_4,f_5)$.
	\EndIf
\end{algorithmic}
\end{figure}

\subsection{Slim method}\label{Subsec:Slim}

The Sparse low-degree implicitization\cite{Slim2005}, or Slim, is an algorithm generating an ball-based local implicit representation.
In other terms, the areas $A_i$ are balls and intersections of balls.
The local approximation procedures $F_i$ are bivariate quadratic polynomials in local coordinates, much like the procedures of type~(\ref{Enum:MPUb}) of the MPU method.
We can use a more restricted variety of local approximation procedures because we have a better control over the positioning of the areas.
Indeed, while the MPU areas are all cuboids (or cubes in the rescaled space) partitioning the object's bounding box, here we use spheres that we can centre on the object's surface, with no fear of having remote areas containing only a small portion of the object in its corner.

The drawback is the need for overlapping spheres in order to cover the whole object.
As polynomial continuity can hardly be satisfied in the overlapping areas, and certainly not with low-degree polynomials, another approach is used: in these areas, the polynomials are weighted depending on the point's distances to the centres of the overlapping balls.
Of course, the weights are computed on-the-fly when the ownership of a query point is asked (or the intersection of the object with a query ray must be computed):
only the local quadratic polynomials tied to single balls are stored in the representation.

Also, the query points that are not covered by the spheres may be inside or outside the object.
When asking the ownership of a query point $q$ in this situation, simply search for its nearest neighbour $p$ in $\Pc$ and check the sign of $<q-p, n>$ where $n$ is $p$'s outer normal.
When negative, $q$ is inside the object with a signed distance close to $-\norm{p-q}$.
When positive, $q$ is outside the object with a signed distance close to $\norm{p-q}$.

Slim uses compactly supported Gaussian-like weights:
\begin{equation}\label{Eq:BumpFunction}
G_R(r):=
\begin{cases}
\exp\left(-\cfrac{1}{1-(r/R)^2}\right) & \text{if } r \in (-R, R) \\
0                                                    & \text{otherwise}
\end{cases}
\end{equation}

We first need to cover $\Pc$ by a set of balls of a given radius.
A simple and efficient way to do it is to pick a random point from $\Pc$ as the centre of the first ball and then continue picking random points as the centres of the subsequent balls amongst those that are not yet covered.
This way, the centres of all the balls used in the algorithm are points of $\Pc$.

Given a ball $B=B(c,R)$, a rough estimation of the surface's normal $n$ close to $c$ is obtained as the average of the normals of $\Pc \cap B$.
A local coordinate system $(u,v,n)$ centred on $c$ is used: quadratic polynomials in this local system are of the form~(\ref{Eq:MPUb}).
The best local approximation w.r.t. the ball $B$ is then the quadratic polynomial $F_B$ minimizing the following quantity:
\begin{equation}\label{Eq:SlimF}
\sum_{p \in \Pc \cap B(c,R)} G_R(\norm{p-c}) F_B(p)^2
\end{equation}

Once a local approximation $F_B$ is computed, two rankings are assigned to a radius $\rho$:
\begin{equation}\label{Eq:SlimRho}
\begin{split}
\epsilon(\rho) :=& \sum_{p \in \Pc \cap B(c,\rho)} F_B(p)^2 \\
E(\rho) :=& \epsilon(\rho) + \lambda (T_{\text{MDL}}/\rho)^2
\end{split}
\end{equation}
where $T_{\text{MDL}}$ is a parameter and $\lambda$ is a regularizing constant computed once:
it is set as the average of the minimum eigenvalues of the co-variance matrices of each point $p\in\Pc$ with its ten nearest neighbours in $\Pc\backslash\{p\}$.

With this, the Slim algorithm consists of the computations of local approximations w.r.t. balls of gradually smaller radius $\rho_k$ and stop when the quantities $E(\rho_k)$ attains a suitable local minimum.
The balls and local approximations computed at each step can be kept in order to have a multi-scale approximation: if only a rough approximation is required for a specific query, we can use the few big balls of early steps instead of the many small balls of late steps.

\begin{figure}
\caption{Algorithm - Slim}
\label{Algo:Slim}
\begin{tabular}{ll}
\textbf{Input}: 	& A point cloud $\Pc$. \\
				& The normals $\Nc$ of that point cloud. \\
\textbf{Output}:	& A suitable local implicit representation for $\Pc$.
\end{tabular}
\begin{algorithmic}[1]
	\State Let $\text{Rep} = \emptyset$.
	\State Let $\Bc_0 = \{B_{00}, B_{01}, \dots\}$ be a cover of $\Pc$ by balls $B_{0i}$ of radius $\rho_0$.
	\CommentTwo{typically,}{$\rho_0$ is 1/10 of the main diagonal of the whole object's bounding box}
	\State Let $U=\Pc$, a list of ``uncovered'' points.
	\State Let $k=1$ and $\rho_1 = g\rho_0$. \Comment{typically, $g=\frac{\sqrt{5}-1}{2}$, the golden ratio conjugate}
	\While{$U$ is not empty}
		\State Let $\rho_{k+1} = g\rho_k$.
		\State Let $\Bc_k = \{B_{k0}, B_{k1}, \dots\}$ be a cover of $U$ by balls of radius $\rho_k$.
		\For{$B\in\Bc_k$}
			\State Compute an approximation $F_B$ by minimizing (\ref{Eq:SlimF}).
			\CommentTwo{typically,}{$T_{\text{MDL}} = 0.02$}
			\If{$E(\rho_{k+1}) > E(\rho_k) < E(\rho_{k-1})$ and $\epsilon(\rho_{k+1}) < \epsilon(\rho_k) < \epsilon(\rho_{k-1})$}
			\linebreak[4]\Comment{see (\ref{Eq:SlimRho})}
				\State Remove the points $\Pc \cap B$ from $U$ and add $(B, F_B)$ to $\text{Rep}$.
			\Else
				\State Optionally store $(B, F_B)$ at the level $k$ of a multi-scale representation.
			\EndIf
		\EndFor
		\State Increment $k$.
	\EndWhile
	\Return the representation $\text{Rep}$.
\end{algorithmic}
\end{figure}

Once the representation structure is generated (see~\ref{Algo:Slim}), the only thing left is how overlapping areas must be dealt with.
Consider a query point $q\in B_1 \cap \dots \cap B_m$ where $B_j$ are balls of centre $c_j$ and radius $r_j$ given by a Slim representation.
Then $F_{B_1 \cap \dots \cap B_m}(q) := \cfrac{\sum_j G_{r_j}(\norm{q-c_j}) F_{B_j}(q)}{\sum_j G_{r_j}(\norm{q-c_j})}$.
The query point $q$ belongs to the object iff $F_{B_1 \cap \dots \cap B_m}(q) \le 0$.
That way, the transitions of the surfaces between the balls $B_j$ are smoothened.

Similarly, the intersection of the object with a query ray $\ell$ is given by $\cfrac{\sum_j G_{r_j}(\norm{q_j-c_j}) q_j}{\sum_j G_{r_j}(\norm{q_j-c_j})}$ where $q_j$ are the intersections of $\ell$ with the local surfaces given by $F_{B_j}$ and the balls $B_j$ taken into account are only the first ones:\\
{\setlength{\tabcolsep}{2pt}
\begin{tabular}{rl}
$\{B_j\}_j :=$ & $\{\text{ball } B \text{ of the Slim representation} \setst \ell \cap B_1 \cap B \ne \emptyset,$ \\
                      & $\text{ where } B_1 \text{ is the first ball intersected by }\ell \}$ (see Fig.~\ref{Fig:SlimRaySmoothing}).
\end{tabular}}

\begin{figure}[tb]
\centering
\includegraphics[width=220pt]{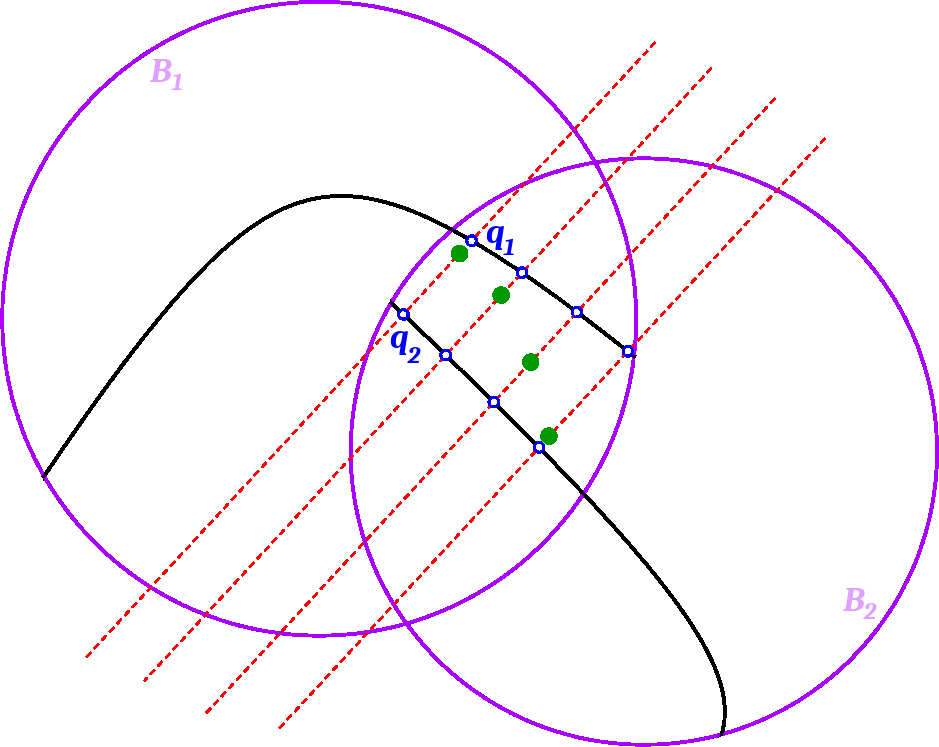}
\caption{Slim: ray intersection in overlapping spheres.}
\label{Fig:SlimRaySmoothing}
\end{figure}

\section{Swept volume data structure}\label{Sec:SweptVolumeDataStructure}

\begin{definition}\label{Def:RigidTransformation}
A \emph{rigid transformation} $T$ is a map

\begin{tabular}{RCCL}
\Tc: & [a, b] & \rightarrow & \Iso(\RR^3) \\
& t & \mapsto & \Translation_{v(t)} \circ \Rotation_{\alpha(t), \beta(t), \gamma(t)}
\end{tabular}
where $v:[a, b] \rightarrow \RR^3$ and $\alpha, \beta, \gamma$ are piecewise polynomials
and $\Translation_v, \Rotation_{\alpha, \beta, \gamma}$ are respectively the translation of vector $v$ 
and the rotation of Euler angles $(\alpha, \beta, \gamma)$.

A \emph{swept volume} $\Tc(\Bc)$ of base $\Bc$ and of rigid transformation $\Tc$ is \linebreak[4]
$\Tc(\Bc):=\cup_{t\in [a, b]} [\Tc(t)](\Bc)$.
\end{definition}

\begin{example}\label{Ex:SweptCapsuleShape}
Let $\Bc$ be a capsule-like shape:
\begin{align*}
\Bc = (& (B((-2, 0, 0), \sqrt{2}), y^2 + z^2 - x - 2),\\
& (B((0, 0, 0), \sqrt{2}), y^2 + z^2 - 1),\\
& (B((2, 0, 0), \sqrt{2}), y^2 + z^2 + x - 2)) \text{ where }B(x, r)\text{ is the ball of centre $x$ and radius $r$}
\end{align*}
And $\Tc$ a linear interpolation between $\Id$ and $\Translation_{(0, 16, 0)} \circ \Rotation_{0, \pi, 0}$:
\begin{align*}
[\Tc(t)](x, y, z) =
\begin{pmatrix}
\cos(\pi t) & 0 & -\sin(\pi t)\\
0 & 1 & 0\\
\sin(\pi t) & 0 & \cos(\pi t)\\
\end{pmatrix} . \begin{pmatrix} x \\ y \\ z \end{pmatrix} + \begin{pmatrix} 0 \\ 16 t \\ 0 \end{pmatrix},
\text{ for } t\in [0,1]
\end{align*}

This swept capsule-like shape is drawn in figure~\ref{Fig:SweptRepAlternatives}.
\end{example}

We describe how, given $\Bc$ and $\Tc$, we construct a local implicit representation of $\Tc(\Bc)$.


In previous works, such implicit representation of swept volumes have been developed for specific types of base volumes.
For instance, the boundary of swept volumes of convex polyhedrons are ruled surfaces; that property is used for the implicitization algorithms described in~\cite{Manocha2004,ZhangKimManocha2009}.
As swept volumes have many applications in robotics and collision detection, another algorithm described in~\cite{Taubig2011} handles base volumes made of shifted convex polyhedrons (i.e. points at a given ``safety'' distance of a convex polyhedron).
Also, in~\cite{Perrin2012}, swept cuboids are approximated for the purpose of a real-time planning of a walking robot's movements.
In the following, though, we assume that the base volume can be anything in the range of the definition~\ref{Def:BaseVolume}.

\begin{figure}[tb]
\hspace{-60pt}
\begin{tikzpicture}[->, >=stealth, x=1pt, y=1pt]
\node (Structure) at (220,0) {
	\includegraphics[width=340pt]{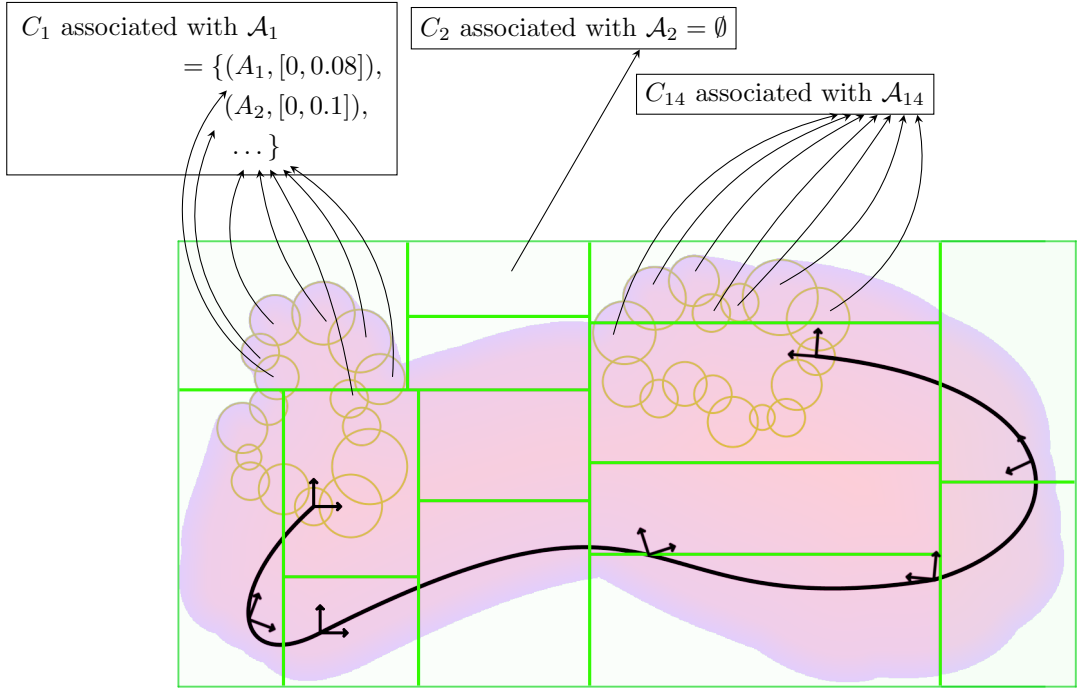}
};

\node (Corner) at (0, 0) {\hphantom{0pt}};
\node (CStart) at (40, 165) {$C_1$ associated with $\mathcal{A}_1$};
\node (CStartA1) at (90, 150) {$= \{ (A_1, [0, 0.08]), $};
\node (CStartA2) at (95, 135) {$(A_2, [0, 0.1]), $};
\node (CStartEtc) at (80, 120) {$\dots \}$};
\draw ($(CStart.north west)+(-2,2)$) rectangle ($(CStartA1.south east)+(2,-32)$);

\node[draw] (CMid) at (200, 165) {$C_2$ associated with $\mathcal{A}_2 = \emptyset$};

\node[draw] (CEnd) at (280, 140) {$C_{14}$ associated with $\mathcal{A}_{14}$};

\draw ($(Structure.north west)+(40,-55)$) to[bend left=50] ($(CStartA1.south west)+(20,0)$);
\draw ($(Structure.north west)+(35,-48)$) to[bend left=35] (CStartA2.south west);
\draw ($(Structure.north west)+(40,-35)$) to[bend left=35] (CStartEtc);
\draw ($(Structure.north west)+(60,-34)$) to[bend left=15] (CStartEtc);
\draw ($(Structure.north west)+(75,-40)$) to[bend right=20] (CStartEtc);
\draw ($(Structure.north west)+(85,-55)$) to[bend right=30] (CStartEtc);
\draw ($(Structure.north west)+(70,-62)$) to[bend right=10] (CStartEtc);

\draw ($(Structure.north west)+(130,-15)$) to ($(CMid.south)+(25,0)$);

\draw ($(Structure.north)+(-5,-39)$) to[bend left=30] ($(CEnd.south)+(20,0)$);
\draw ($(Structure.north)+(10,-20)$) to[bend left=20] ($(CEnd.south)+(25,0)$);
\draw ($(Structure.north)+(26,-15)$) to[bend left=15] ($(CEnd.south)+(30,0)$);
\draw ($(Structure.north)+(33,-30)$) to[bend left=8] ($(CEnd.south)+(35,0)$);
\draw ($(Structure.north)+(42,-28)$) to[bend right=5] ($(CEnd.south)+(40,0)$);
\draw ($(Structure.north)+(58,-20)$) to[bend right=25] ($(CEnd.south)+(45,0)$);
\draw ($(Structure.north)+(77,-30)$) to[bend right=35] ($(CEnd.south)+(50,0)$);

\end{tikzpicture}
\caption{Tree structure $(C_j, \mathcal{A}_j)_j$ in 2D with circles as local areas (\emph{yellow}). The rigid transformation (\emph{black} curve with orientation) is applied on the local areas (\emph{purple} surface) and used to construct the cells (\emph{green}): each one of these cells is associated with the part of $\Bc$ and the time span that are relevant. Note that the local implicit procedures $F_i$ are not involved at this step.}
\label{Fig:SweptStructure}
\end{figure}

Let $\Bound$ be a bounding box of $\Tc(\Bc)$.
We split $\Bound$ into cells $(C_j)_{1 \le j \le M}$ and compute \linebreak[4]
$\mathcal{A}_j := \left\lbrace (A_i, [t_0, t_1]) \mid \forall t\in [t_0, t_1], C_j \cap [\Tc(t)](A_i) \ne \emptyset \right\rbrace$.
It is the list of local areas of $\Bc$ intersecting the cell $C_j$ along the swept transformation and the times between which they intersect (see the figure~\ref{Fig:SweptStructure}.
How we split $\Bound$ into cells and how we compute $(\mathcal{A}_j)_j$ in practice is explained further.

Given a swept volume, we choose a suitable partition $(C_j)_j$ and compute $(\mathcal{A}_j)_j$ once.
The tree structure given by $(C_j, \mathcal{A}_j)_j$ is our preprocessing structure.
It allows to filter the relevant areas used for checking whether a point belongs to the swept volume or not.
Proceeding that way, the local procedures $F_i$ are not requested at all at the preprocessing step: only the intersection of relatively simple objects, the moving areas (moving spheres, moving cuboids,\dots) and the cells (rectangular cuboids), must be computed.

Once the tree structure is known, let $P\in\RR^3$ be a query point.
If $P\not\in\Bound$, we return that $P\not\in\Tc(\Bc)$.
Else, using the preprocessing structure, we find $j$ such that $P\in C_j$ in $O(\log(M))$ time complexity.
We then perform more accurate checks on $P$, using a numerical solver to find
\[
\min \left\lbrace F_i([\Tc(t)^{-1}](P)) \mid A_i \text{ and } t \text{ are in }\mathcal{A}_j\right\rbrace
\]
This can be performed in $O(|\mathcal{A}_j| \log(\epsilon^{-1}) \tau_j)$ worst-time complexity using the bisection algorithm, where $\epsilon$ is the solver precision and $\tau_j$ the size of the time segments $[t_0, t_1]$ in $\mathcal{A}_j$.
It can be performed faster if the hypotheses on $F_i$ allow better algorithms to be used (typically, the Newton method when one can compute the differential of $F_i$).

Thus, we want $|\mathcal{A}_j|$ and $\tau_j$ to be rather small.
We are interested in computing a partition of $\Bound$ by cells $(C_j)_j$ minimizing the following quantity:
\begin{equation}\label{Eq:SweptCost}
\Cost((C_j)_j) := \log(M) + \frac{1}{M} \sum_{j=1}^M \Vol(C_j) |\mathcal{A}_j| \tau_j
\end{equation}

The use of a mean measure weighted by the size of cells instead of the maximal value is motivated by the objective to give an implicit procedure that would likely be used on a lot of points.
One can add more sophisticated weights if parts of the models are more likely to be processed than others (for instance, if there is a visible face of the swept volume and a back face that is not usually rendered).
Such a weight can be introduced by considering
$\Cost_\omega((C_j)_j) := \log(M) + \frac{1}{\int_{x\in \Bound}\omega(x) \diff x}\sum_{j=1}^M \left(\int_{x\in C_j}\omega(x) \diff x |\mathcal{A}_j|\tau_j\right)$
where $\omega: \Bound \rightarrow \RR_+$ is a bounded user-specified weight that is high-valued in the important areas of the swept volume and low-valued in less important areas.

In order to minimize the cost, we split the bounding box of the swept volume, $\Bound$, according to the following procedure:
\begin{enumerate}
\item Start with a trivial partition $C_1 := \Bound$.
\item Pick many parameters $(t_k)_k$ in $[a, b]$ and consider the rigid transformation at time $t_k$ applied to the local areas, $S_{i,k} := [\Tc(t_k)](A_i)$.
\item While the cost of the partition~(\ref{Eq:SweptCost}) decreases, pick the cell with the largest (weighted) volume and split it along a coordinate in two other cells $c_1, c_2$ by optimizing\linebreak[4]
$\#\left\{ (i,k) \mid S_{i,k} \cap c_1 \ne \emptyset \right\} + \#\left\{ (i,k) \mid S_{i,k} \cap c_2 \ne \emptyset \right\}$.
\item For the cells of the boundary, find the best split that would generate an emply cell (i.e. with no intersection with $\cup_{i,k} S_{i,k}$).
If that empty cell has a surface large enough (possibly weighted by $\omega$), then perform the split.
\end{enumerate}


Now, we develop the way to compute $\mathcal{A}_j$, the local areas intersecting the cell $C_j$.
This step relies heavily on the basic shapes used for the local areas $A_i$;
the method must be adapted depending on what shape is used.
Since the cells $C_j$ themselves are rectangular cuboids, the computation of $\mathcal{A}_j$ consists of solving rectangular cuboid/rectangular cuboid intersection problems (when $\Bc$ was generated by MPU) or sphere/rectangular cuboid intersection problems (when $\Bc$ was generated by Slim) etc., one of which being moving (i.e. depending on a parameter $t$).
Either $A_i$ or $C_j$ can be chosen to depend on the time parameter; this choice corresponds to solving either one of the two equivalent problems:
\begin{align}
& \text{Solve } [\Tc(t)](A_i) \cap C_j \neq \emptyset \text{ w.r.t. $t$},\label{Eq:MovingShapeCollusion1} \\
& \text{Solve } [\Tc(t)^{-1}](C_j) \cap A_i \neq \emptyset \text{ w.r.t. $t$}.\label{Eq:MovingShapeCollusion2}
\end{align}
When $A_i$ is a sphere, it is more efficient to use the first alternative since it means applying $\Tc(t)$ less times (we apply it only on the centre of the sphere, instead of applying it to each of the 6~cuboid's faces).
When $A_i$ is a more complicated shape than $C_j$, we use the second alternative instead.
That is what we do when deciding whether a point $P$ belongs to $\Tc(\Bc)$: we compute parametrically the ownership of $[\Tc(t)^{-1}](P)$ to $\Bc$ instead of the ownership of $P$ to $[\Tc(t)](\Bc)$.

Let $f_{k,-1}, f_{k,+1}$ (with $k\in\{1,2,3\}$) be the normalised equations of the 6 faces of $C_j$.
For ease of notations, we will use $k$, $k'$ and $k''$ such that $\{ k, k', k'' \} = \{ 1, 2, 3 \}$ so that each one corresponds to one coordinate.
By \emph{normalised equations} of faces, we mean equations of the form $f_{k,\pm1}(P) = P.\vec{n} - d$ where $\vec{n}$ is the unit outward-pointing normal and $d$ is a suitable constant ($d=Q.\vec{n}$ for a point $Q$ of the face).
This way, $f_{k,\pm1}$ are the signed-distance functions of the faces of $C_j$.
Notice that, since the cell $C_j$ has the same orientation as the axes, $f_{k,\pm1}$ actually depends only on one coordinate.
Also, let $e_{k,\sigma_1,k',\sigma_2}$ (with $\sigma_k \in \{ -1, +1 \}$) be the edge of $C_j$ defined by $f_{k,\sigma_1}=f_{k',\sigma_2}=0$ and $v_{\sigma_1,\sigma_2,\sigma_3}$ be the vertex defined by $f_{1,\sigma_1}=f_{2,\sigma_2}=f_{3,\sigma_3}=0$.

Now, suppose that $A_i$ is a sphere of centre $O$ and radius $R$.
The moving centre $[\Tc(t)](O)$ is thus given by $O(t) := M(t).O + v(t)$ where $M(t)$ is the rotation matrix of Euler angles $(\alpha(t),\beta(t),\gamma(t))$.
The problem~(\ref{Eq:MovingShapeCollusion1}) can then be described by the following equations:
\begin{align*}
& f_{k,\sigma_1}(O(t))-R_i = 0 \setand f_{k',\pm1}(O(t)) \le 0 \setand f_{k'',\pm1}(O(t)) \le 0 \\
\setor & \begin{tabular}{l}
$\Dist(O(t), e_{k,\sigma_1,k',\sigma_2})^2 - R^2 = 0 \setand f_{k,\sigma_1}(O(t)) > 0$ \\
$\setand f_{k',\sigma_2}(O(t)) > 0 \setand f_{k'',\pm1}(O(t)) \le 0$
\end{tabular}\\
\setor & \begin{tabular}{l}
$\Dist(O(t), v_{\sigma_1,\sigma_2,\sigma_3})^2 - R^2 = 0 \setand f_{1,\sigma_1}(O(t)) > 0$ \\
$\setand f_{2,\sigma_2}(O(t)) > 0 \setand f_{3,\sigma_3}(O(t)) > 0$
\end{tabular}
\end{align*}
which makes 6~equations to solve for the first case, plus 12 for the second case and 8 for the third case for a total of 26~equations per sphere/cuboid couples.
For all the solutions found, several inequalities must be checked but these are not expensive.

\begin{remark}\label{Rem:SpeedUpSweptPreprocess}
Note that it is possible to approximate the structure $\mathcal{A}_j$ by solving $f_{k,\sigma_1}(O(t))-R_i = 0$ instead.
When doing that, there are only 6~equations to solve per sphere/cuboid couples, which effectively makes the preprocessing computation faster at the price of a slightly slower runtime for membership checks and ray intersections.

An other way to speed up this preprocessing step, notice that if an area $A_i$ is in contact with a cell $C_j$ for $t\in[t_0, t_1]$, then the $A_{i'}$ cannot be in contact with any cell $C_{j'}$ such that $\Dist(C_j, C_{j'}) + \Diameter(A_i) + \Diameter(A_{i'}) > \Dist(A_i, A_{i'})$ in the same time period.
Thus, using the informations on the already computed area positions allows to filter out a few cells when processing the areas that are nearby the former one.
\end{remark}


The algorithm~\ref{Algo:SweptVolume} sketches how an implicit representation of $\Tc(\Bc)$ is computed and the algorithm~\ref{Algo:SweptVolumeUsage} shows how to use that implicit representation, both as an ownership oracle and as a ray intersection test.

\begin{figure}
\caption{Algorithm - Implicit representation of swept volume from implicit representation of base volume}
\label{Algo:SweptVolume}
\begin{tabular}{ll}
\textbf{Input}: 	& A base volume $\Bc$, possibly given with distance functions. \\
				& A rigid transformation $\Tc$. \\
\textbf{Output}:	& A procedural implicit representation of $\Tc(\Bc)$, possibly allowing distance computation.
\end{tabular}
\begin{algorithmic}[1]
	\Statex{\texttt{/* \mbox{Computing a bounding box can be done by $\min_x := r + \min_t(v(t)_x)$,} etc. where $r$ is the radius of a bounding sphere of $\Bc$ and $v$ is the translation vector of $\Tc$ */}}
	\State Compute a bounding box $\Bound$ of $\Tc(\Bc)$.
	\Statex{\texttt{/* Compute a suitable partition of $\Bound$ */}}
	\State Let $C := \{\Bound\}$
	\For{$i = 0,\dots, M$}
		\State Split $C$ at the position $[\Tc(a+i(b-a)/M)](\Center(\Bc))$ w.r.t. the coordinate maximizing $|v'(a+i(b-a)/M)|$
	\EndFor
	\Statex{\texttt{/* Setup the tree structure of the representation */}}
	\For{$c \in C$}\Comment{see the remark~\ref{Rem:SpeedUpSweptPreprocess} for a smarter loop}
		\Statex{\texttt{/* Computing the intersection of a moving area with a cuboid.}}
		\Statex{\texttt{   The formulae depend on the type of the moving area (cube, sphere\dots).*/}}
		\State Compute the local areas $\mathcal{A}_c := \left\lbrace (A_i, [t_0, t_1]) \mid \forall t\in [t_0, t_1], c \cap [\Tc(t)](A_i) \ne \emptyset \right\rbrace$
	\EndFor
	\Return $C, \{\mathcal{A}_c\}$
\end{algorithmic}
\end{figure}

\begin{figure}
\caption{Algorithm - Usage of the swept volume implicit representation provided by the algorithm~\ref{Algo:SweptVolume}}
\label{Algo:SweptVolumeUsage}
\begin{tabular}{ll}
\textbf{Usage}: 	& Check ownership of a query point $P$. \\
				& Compute intersections with a query ray $R$.
\end{tabular}
\begin{algorithmic}[1]
	\Statex{\texttt{/* Ownership of $P$ */}}
	\State Find cell $c\in C$ such that $P \in c$
	\If{there is no such cell}
		\Return false, ``$P$ is far away''
	\EndIf
	\State Let $d \gets +\infty$
	\ForAll{$(A_i, [t_0, t_1]) \in \mathcal{A}_c$}
		\State Solve $(t, d_{\text{tmp}}) \gets \min_{t\in[t_0, t_1]}(F_i([\Tc(t)]^{-1}(P)))$ (*)
		\If{$[\Tc(t)]^{-1}(P)$ is on the inner boundary of $A_i$}
			\Return true, ``$P$ is far inside''
		\EndIf
		\State Let $d \gets \min(d, d_{\text{tmp}})$
	\EndFor
	\Return $d \le 0, d$ \CommentTwo{$d$ is a signed distance of $\Tc(\Bc)$,}{assuming $F_i$ are local signed distances of $\Bc$}
\end{algorithmic}
\hrule
\begin{algorithmic}[1]
	\Statex{\texttt{/* Intersection with ray $R = \{R_o + s R_d \mid s \in \RR_+ \}$ */}}
	\State Find cells $C_R$  such that $R \cap c \neq \emptyset, \forall c \in C_R$
	\State Sort $C_R$ by distance w.r.t. $R_o$
	\ForAll{$c \in C_R$}
		\ForAll{$(A_i, [t_0, t_1]) \in \mathcal{A}_c$}
			\State Let $I \gets \{t \in [t_0, t_1] \mid [\Tc(t)]^{-1}(R) \cap A_i \neq \emptyset \}$ (**)
			\If{$I \ne \emptyset$}
				\State \mbox{Let $s \gets \min(\{s \mid t \in I, [\Tc(t)]^{-1}(R(s))\in A_i, F_i([\Tc(t)]^{-1}(R(s))) \le 0\})$ (*)}
				\Return $R(s)$
			\EndIf
		\EndFor
	\EndFor
	\Return ``$R$ does not intersect $\Tc(\Bc)$''
\end{algorithmic}
(*) Using Newton or bisection algorithms depending on the properties of $F_i$\\
(**) Using the Newton algorithm if a suitable ray/object distance is provided or the bisection algorithm else
\end{figure}

\clearpage

\bibliographystyle{plain}

\end{document}